 \documentclass{aa}
\usepackage{graphics}
\newcommand {\vmi}   {$v_{\mathrm {micro}}$}
\newcommand {\vma}   {$v_{\mathrm {macro}}$}

\newcommand {\Tef}   {$T_{\mathrm {eff}}$}
\newcommand {\K}     {{$\mathrm{~K}$}}
\newcommand {\lgg}   {$\mathrm \log ~g$}

\newcommand {\kms}   {${\mathrm {km~s^{-1}}}$}
\newcommand {\hd}    {~HD221170~}

\begin{document}

\title{Thorium-rich halo star HD221170: further evidence against
 the universality of the  r-process$^{*}$}

 \author{Alexander Yushchenko	  \inst{1,2}	      \and
  Vera  Gopka                     \inst{2}            \and
  Stephane Goriely		  \inst{3}	      \and
  Faig  Musaev                    \inst{4, 5, 6  %, 10
                                               }    \and
  Angelina Shavrina		  \inst{7}	      \and
  Chulhee Kim                     \inst{8}            \and
  Young Woon Kang,                \inst{1}            \and
  Juliana Kuznietsova,            \inst{5,7}          \and
  Vladimir Yushchenko             \inst{9}
     }
\offprints{A.V. Yushchenko \\
		 $^{*}$ Based on observations obtained at the 2-m telescope
			of Peak Terskol observatory,
		      % near Mt. Elbrus,
		      % Northern Caucasus, Russia --
			International Center for
			Astronomical, Medical and Ecological Research,
				  Ukraine \& Russia }

\institute{
     Astrophysical Research  Center for the Structure and Evolution        %1
     of the  Cosmos (ARCSEC),  Sejong University, Seoul, 143-747, Korea
     \email{yua@arcsec.sejong.ac.kr, kangyw@sejong.ac.kr}
\and
        Odessa Astronomical observatory, Odessa National University,       %2
	Park Shevchenko, Odessa, 65014, Ukraine
     \email{yua@odessa.net, gopka@arctur.tenet.odessa.ua}
\and
        Institut   d'Astronomie  et  d'Astrophysique,   Universite  Libre  %3
	de Bruxelles, CP 226, 1050 Brussels, Belgium
	\email{sgoriely@astro.ulb.ac.be}
\and
     Special Astrophysical observatory of the Russian Academy of Sciences, %4
     Nizhnij Arkhyz, Zelenchuk, Karachaevo-Cherkesiya, 369167, Russia
     \email{faig@sao.ru, gala@sao.ru}
\and
     The International Centre for Astronomical, Medical and Ecological     %5
     Research of the Russian Academy of Sciences and
     the National Academy of Sciences of Ukraine,
     Golosiiv, Kiev, 03680, Ukraine % (ICAMER)
     \email{admin@terskol.com}
\and
     Shamakhy   Astrophysical   Observatory,	NAS   of  Azerbaijan,  Yusif %6
     Mamedaliev, Shamakhy, Azerbaijan
\and
     Main Astronomical observatory, NAS of Ukraine, Kiiv, 03680, Ukraine   %7
     \email{shavrina@mao.kiev.ua}
\and
     Department of Earth Science Education,                                %8
     Chonbuk National University,
     Chonju 561-756, Korea,
     \email{chkim@astro.chonbuk.ac.kr}
\and
     Odessa National University, Odessa, 65014, Ukraine                    %9
     \email{yushchenko@mail.ru}
}
\date{Received March XX, 2003; accepted Xxxxx XX, 2003}

 \abstract{
       We  report  the abundance determination  in the atmosphere of the
  bright  halo  star HD221170. The spectra  were  taken with the Terskol
  Observatory's   2.0-m   telescope  with   a   resolution  R=45000  and
  signal-to-noise ratio up to 250 in the wavelength region 3638-10275 \AA.~
  The   adopted  atmospheric  parameters   correspond  to  an  effective
  temperature  \Tef=4475~K, a surface gravity \lgg=1.0, a microturbulent
  velocity \vmi=1.7 \kms, and a macroturbulent velocity \vma=4 \kms. The
  abundances  of 43 chemical elements were determined with the method of
  spectrum synthesis.
%  The abundances of S, Mo, Er, Tm, Hf,
%  W, Os, Ir, Pb, Th, and upper limit for U were found for
%  the first time.
  The  large overabundances (by 1 dex relative to iron) of elements with
  Z$>38$ are shown to follow the same pattern as the solar r-abundances.
  The  present  HD221170 analysis confirms  the  non-universality of the
  r-process,  or  more  exactly the  observation  that the astrophysical
  sites  hosting  the  r-process  do not  always  lead  to a unique
  relative  abundance  distribution for the bulk  Ba to Hg elements,
% on a first side,
  the Pb-peak elements,
% on a second side
  and the actinides.
% on a third side.

 %
 %
 %
}

\titlerunning{Thorium-rich halo star HD221170}

 \maketitle

\section{Introduction}

  The  chemical  composition  of  different  objects,  particularly halo
  stars,  is  one  of  the most  important  clues  for understanding the
  structure and evolution of the Universe. Typical investigations of the
  chemical  compositions of stars usually  show 15-40 chemical elements,
  depending
% of
  the type of the  star.
  The most detailed stellar abundance
  patterns  consist of  50-58 elements (see Yushchenko et
  al.  2004 for examples).
  The best  example is the abundance pattern of
  the halo star CS22892-52 with the determination of 58 elements (Sneden
  et  al.  2003).  This  star  and  another  three  r-process-rich stars
  CS31082-001  (Cayrel  et al. 2001,
% {\bf
   Hill et al. 2002
%  }
   ),
   HD115444  (Westin et al. 2000) and
  BD+17$^o$3248  (Cowan  et  al.  2002) are
% the
  halo  stars with known
  enhanced  abundances of thorium with respect to iron. In the specific
  case  of CS31082-001, the determination  of the uranium abundance also
  enabled
  investigators
  to  derive  from  the thorium  to  uranium  abundance ratio a
  stellar  age  of  13$\pm$4 billion  years.  Such  an estimate directly
  provides  an independent lower limit for the age of our Galaxy and the
  Universe.

% It should be noted that
  The investigation of the thorium abundance in
  the  atmosphere  of different stars of  our Galaxy has a long history.
  Bucher  (1987), Morel et al. (1992) and Francois et al. (1993) studied
  a  set  of  disk  and  halo  stars  using  the  strongest  Th  line at
  4019.129~\AA.  Yushchenko  \&  Gopka  (1994)  determined  the  thorium
  abundance  in  Procyon  using  4 faint  lines  in  the  3200-3500 \AA~
  spectral range, while Gopka et al. (1999) determined Th at the surface
  of Arcturus using other lines.

  A  special  effort  was  later  dedicated  to  the  observation  of an
  increasing  number  of Th-rich halo stars  (see for example Johnson \&
  Bolte,  2001).  But the number of  lines  considered for the abundance
  determination  did  not necessarily increase;  for  example the latest
  investigation  of Honda et al. (2004) is based on the single 4019 \AA~
  line.   Numerous  thorium  lines  were   detected  only  in  the  four
  above-mentioned Th-rich stars.

  Many authors have used the Th/Eu %,  Th/U
  and other ratios to estimate the age
  of  the observed stars, although Goriely  \& Clerbaux (1999) and Goriely
  \&  Arnould argued that the Th-to-Eu abundance ratio is not a reliable
  chronometer  to  derive the stellar age.  New observations by Honda et
  al. (2004), as well as the present analysis confirm this result.

  \hd  is  one of the bright  template  halo stars. This seven-magnitude
  star has been investigated regularly as  new methods and new observational
  facilities  became  available.  The  first  paper  in  this series was
  Wallerstein  et al. (1963), followed by Gilroy (1988) and many others.
  The last detailed abundance pattern of heavy elements in this star was
  provided  by Burris et al. (2000).  Yushchenko et al. (2002) published
  preliminary  results  for the abundances  of elements heavier than Dy.
  However,  for  such  elements the majority  of  the spectral lines are
  located   at   wavelengths   shorter  than   4500~\AA,   and  the  low
  signal-to-noise  ratio  obtained in this  spectral region at that time
  required   further  investigations.  Gopka  et  al.  (2004;  hereafter
  Paper~I)  used the spectral data from the above-mentioned work and the
  spectrum  from  the archive of  the  Haute-Provence Observatory 1.88 m
  telescope to derive the abundance of elements with $Z\le 68$. Previous
  investigations  concerning  the  surface  abundances  of  \hd  and the
  determination  of  atmospheric  parameters  are  reviewed  in Paper~I.
  According  to  Burris et al. (2000),  Gopka et al. (2001) and Paper~I,
  the  iron abundance in \hd is deficient by about 2 dex with respect to
  the  sun  while  the  abundance of  elements  heavier  than barium are
  overabundant  with  respect to iron.  Nevertheless, since the spectrum
  used in Paper~I from the Haute-Provence Observatory covers the limited
  wavelength  range  of 4480--6820 \AA, we  were unable to determine the
  abundance of elements heavier than Dy.

  In  Sect.~2,    new  observations  of  higher  quality  and  broader
  wavelength   coverage   are  detailed   and  the  adopted  atmospheric
  parameters  described. In Sect.~3 the result of the abundance analysis
  is  presented.  In  Sect.~4, the  surface  abundances  of HD221170 are
  compared with those of the other r-process-enriched halo stars already
  observed, and their implication concerning the non-universality of the
  r-process is discussed.

\section{Observations and atmospheric parameters }

  The  observations  were conducted  with the Coude-echelle spectrometer
  (Musaev  et al. 1999) mounted at the 2-m "Zeiss" telescope at the Peak
  Terskol  observatory  located  near  Mt.  Elbrus  (Northern  Caucasus,
  Russia)  3,124 m above sea level. The spectrometer is used in the mode
  with  a spectral resolution of  45,000. The best signal-to-noise ratio
  is  over  250 in the red  spectral  region and the wavelength coverage
  includes  the  3638-10275~\AA~~ range.  Galazutdinov (1992) DECH20 and
  Yushchenko (1998) URAN codes are used for the data processing.

  In  Paper  I,  based on previous  investigations  of  the star and new
  spectral  data, the following model  atmosphere parameters of \hd were
  found   to   be   optimum:   \Tef=4475$\pm$50\K   ~for  the  effective
  temperature,  \lgg=1.0$\pm$0.1 for  the gravity, \vmi=1.7$\pm$0.1~\kms
  ~for   the   microturbulence   and   \vma=4.0$\pm$0.5~\kms   ~for  the
  macroturbulence.

  These parameters and the abundances  from Paper~I
  are
  used  to  construct  individual atmosphere  models  using the standard
  Kurucz's   ATLAS12  code.  We  find  for  the  final  model  that  the
  correlation(s)  of  iron  abundances  with  excitation  potentials and
  equivalent widths of iron lines are close to zero.

 We also calculated the models corresponding to the parameter sets
 (\Tef=4475\K; \lgg=1.2) and
 (\Tef=4575\K; \lgg=1.0).
 These models are used to estimate the abundance errors due to possible
 uncertainties in the adopted value of the temperature  and gravity.

\section{Abundance analysis}

% {\bf
  The equivalent  widths of the iron lines are estimated by fitting their
  profiles with a  Gaussian function.
%  }
  The  iron abundance is calculated using the model atmosphere method on
  the   basis  of  the  Kurucz   (1995)  WIDTH9  program.
  In  contrast,
  differential  spectrum  synthesis methods are  used  for all the other
  elements. For each line, we tried to find its counterpart in the solar
  spectrum  atlas  of Delbouille et al.  (1973). This procedure frees us
  from  uncertainties  connected with  oscillator  strengths of spectral
  lines.  The URAN code (Yushchenko  1998) and SYNTHE spectrum synthesis
  program (Kurucz 1995) are used to approximate the observed spectrum with
  the  synthetic one. The solar abundances  of Grevesse \& Sauval (1999)
  are  considered. A synthetic spectrum of  \hd for the whole wavelength
  range  helps  us to identify  spectral  lines.  SYNTHE program (Kurucz
  1995)  was used to produce the synthetic spectrum.
  Atomic
  and  molecular  lines
  are included
  from Kurucz (1995)  as  well  as Morton (2000),
  Biemont  et al. (2002) and partially  from the VALD database (Piskunov
  et al.\ 1995)

  Holweger's  partition  function  for thorium  is  used  (Morell et al.
  1992).
  Hyperfine  structure and isotopic splitting are taken into account for
  Li, Sc,  V,  Mn, Co,  Cu,  Ba, and Eu. The  splitting  data  for
%{\bf
   Li are taken from Shavrina et al. (2003),
%   }
  for
  Ba -- from  Francois  (1996)
  and for other elements  from Kurucz (1995).
% It should  be  noted  that
  We found counterparts   in  the  solar  spectrum
  for all elements except Li,  S,  K, Ir, Th, and U,
  so  that  the  differential
  abundances   are   not  strongly   influenced  by  splitting  effects.

% {\bf
  This is true if the lines have approximately the same strength in both
  stars. However, \hd and the Sun are different stars and it is quite difficult
  to find lines of similar intensity. For example, for Hf the equivalent
  widths of two of lines used are of the order of 2 and 3.5 m\AA~ in the Sun
  and near 45 and 15 m\AA~ in \hd. The first line, $\lambda$~3918.094~\AA,
  with an equivalent width of 2 and 45 m\AA~ in the spectra of the Sun and \hd,
  respectively,
  shows a relative abundance difference of hafnium  -0.97 dex
  in the atmosphere
  of \hd with respect to the solar atmosphere.
  In the case of the second line, $\lambda$~4093.155~\AA, the difference
  in the equivalent width is  smaller (3.5 and 15 m\AA) and the relative abundance is
		-1.49 dex.
  We can expect that taking into account the splitting of the lines will
  decrease the overabundance of Hf with respect to iron in the atmosphere
  of \hd. The value obtained using the second line is therefore expected to be more
reliable.
  But even this value shows a non-negligible deviation with respect to the solar
  r-process pattern, as will be shown in the next section of this paper.
%  }

 % Figure 1
 \begin{figure}
 \resizebox{\hsize}{!}{\includegraphics{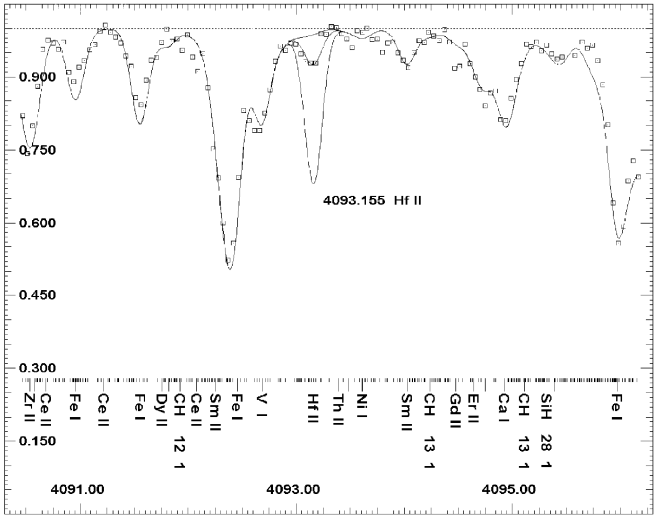}}
 %\vskip 1.5cm
 \caption[]
   {  The  observed spectrum of \hd  (squares) and the synthetic spectra
  (solid  lines) calculated with our final  abundances. The axes are the
  wavelength  in  angstroms  and relative  fluxes.  The positions of the
  spectral  lines  taken into account in  the calculations are marked in
  the  bottom  part  of the figure
  {by short and long dashes.}
  For  some  of  the strong lines, the
  identification and isotopic composition for molecular lines are given.
  The  position of the Hf II 4093.155  \AA~ line is marked by a vertical
  dotted  line.  The  different  synthetic  spectra  correspond  to a Hf
  abundance  lower  or higher by 0.5~dex  with  respect to the abundance
  obtained from the optimum value.}
 \label{fig1}
 \end{figure}

 % Figure 2
 \begin{figure}
 \resizebox{\hsize}{!}{\includegraphics{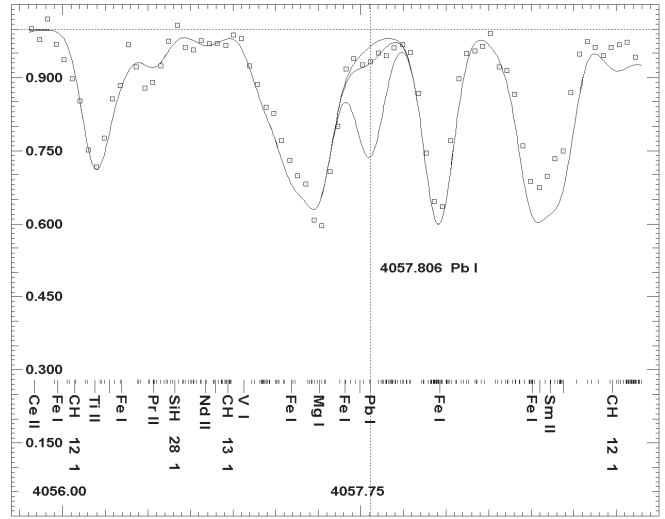}}
 %\vskip 1.5cm
 \caption[]
 {  Same as Fig.~1 in
    the vicinity of the Pb I 4057.806 \AA~ line. The three different
    synthetic spectra correspond to the optimum
    lead  abundance and a deviation by $\pm 0.5$~dex from this value.
%{\bf
  Note that the unmarked strong line on the blueside of the Pb line corresponds
 to the  $^{12}$CH line. The separation of the $^{12}$CH and Pb lines
  is smaller than the instrumental profile, so that the uncertainties in the
  determination of the wavelength
  and oscillator strength can strongly affect the Pb abundance determination.}
%  }
 \label{fig2}
 \end{figure}

 % Figure 3
 \begin{figure}
 \resizebox{\hsize}{!}{\includegraphics{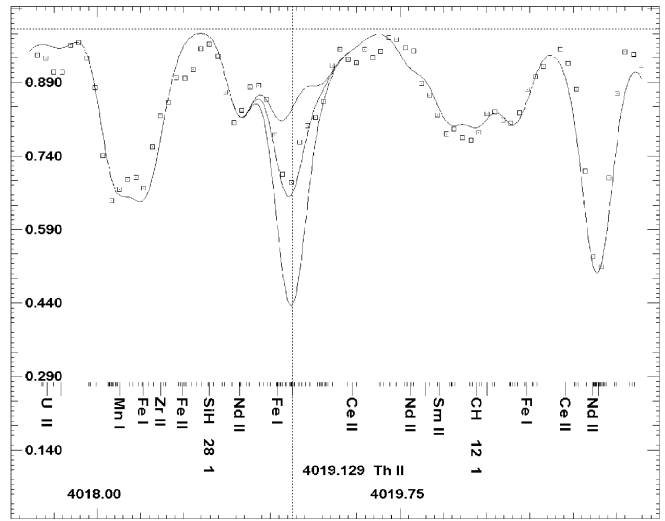}}
 %\vskip 1.5cm
 \caption[]
   {Same  as Fig.~1 in the vicinity of the Th II 4019.129 \AA~ line. The
  different  synthetic spectra correspond to  the mean thorium abundance
  of  -1.18 (in the scale log N(H)=12)  and a deviation by $\pm 0.5$~dex
  from this value.}
 \label{fig3}
 \end{figure}

  For  the  lines  of  several  heavy  elements  (U,  Th,  Os,  Ir),  no
  counterparts  in  the solar spectrum  exist,  and therefore the latest
  values  of  the oscillator strength  (Nilsson et al. 2002a,b, Ivarsson
  2003) are adopted. Several lines of hafnium, lead, thorium and uranium
  can  be  found in Fig.~1--6. The line  data of iron and other chemical
  elements observed in the photosphere of \hd can be obtained in  table
  format
  at      the   websites:
  ``{\tt users.odessa.net/$^\sim$yua}'' and
  ``{\tt yushchenko.netfirms.com}'' as well in electronic Tables 1, 2~
 at the website of this Journal.

\setcounter{table}{2}

\begin{table*}
\caption{
 The mean abundances of chemical elements in the atmosphere of \hd
 with respect to their abundances in the solar atmosphere.
  }
% { \tiny
\begin{tabular}{rrlc lr c lr  c lll r}
 \hline
  &    &         &&\multicolumn{5}{c}{ Paper~I}        &&
                   \multicolumn{4}{c}{ This paper} \\
 \cline{5-9}  \cline{11-14}
  &    &         &&\multicolumn{2}{c}{ Haute-Provence} &&
                                   \multicolumn{2}{c}{Terskol }  &&
                                   \multicolumn{4}{c}{Terskol, new spectrum} \\
 \cline{5-6}  \cline{8-9}  \cline{11-14}
  &~Z~~& Ident.&&$[N/N_H]$ &  n &~& $[N/N_H]$ & n &~&
                                                   \multicolumn{3}{c}{$[N/N_H]$} &      n \\
 \cline{5-6}  \cline{8-9}  \cline{11-13}
  &    &         &&\multicolumn{2}{c}{\rm{ATLAS9}}
                                   &&\multicolumn{2}{c}{\rm{ATLAS9}}
                                                    &&\rm{ATLAS12}
                                                                & \lgg+0.2&\Tef+100\K &     \\
 \cline{1-3}  \cline{5-6}  \cline{8-9}  \cline{11-14}
1 &  3& Li~I    &&          &   &&          &  &&$<$-1.64(20)& $<$-1.64(20)& $<$-1.64(20)&  2  \\
2 &  6& C$^1$   &&          &   &&          &  &&~~-2.7     &           &           &     \\
3 &  7& N$^1$   &&          &   &&          &  &&~~-2.13    &           &           &     \\
4 &  8& O~~I    &&~~-1.86   &  1&&          &  &&~~-1.70(07)&~~-1.66(08)&~~-1.68(12)&  5  \\
5 & 11& Na~I    &&~~-2.43(00)&  2&& -2.22(05)& 2&&~~-2.50(09)&~~-2.51(08)&~~-2.43(11)&  3  \\
6 & 12& Mg~I    &&~~-1.69(08)&  2&& -1.89(05)& 2&&~~-1.76(09)&~~-1.77(09)&~~-1.64(12)&  4  \\
7 & 13& Al~~I   &&$<$-1.73  &  2&&          &  &&$<$-2.11(02)&$<$-1.99(09)&$<$-1.95(05)&  2  \\
8 & 14& Si~~I   &&~~-1.63(08)&  6&& -1.60(05)& 3&&~~-1.77(16)&~~-1.76(17)&~~-1.71(17)& 26  \\
  &   & Si~~II  &&          &   &&          &  &&~~-1.75    &~~-1.73    &~~-1.78    &  1  \\
9 & 16& S~~~I   &&          &   &&          &  &&~~-1.72    &~~-1.64    &~~-1.85    &  1  \\
 10 & 19& K~~I  &&          &   &&          &  &&~~-1.55(11)&~~-1.57(11)&~~-1.34(14)&  2  \\
 11 & 20& Ca~I  &&~~-1.86(07)& 21&& -1.84(15)&10&&~~-1.96(12)&~~-1.98(12)&~~-1.86(11)& 31  \\
  &   & Ca~II   &&          &   &&          &  &&~~-1.67(03)&~~-1.58(03)&~~-1.78(03)&  2  \\
 12 & 21& Sc~I  &&          &   &&          &  &&~~-1.96(09)&~~-1.97(08)&~~-1.81(17)&  3  \\
  &   & Sc~II   &&~~-1.93(08)&  9&& -1.92(00)& 2&&~~-2.12(13)&~~-2.06(13)&~~-2.08(11)& 20  \\
 13 & 22& Ti~I  &&~~-2.00(09)& 43&& -1.97(07)& 8&&~~-2.00(12)&~~-2.01(09)&~~-1.81(09)& 84  \\
  &   & Ti~II   &&~~-1.77(10)& 19&& -1.86(14)& 9&&~~-1.76(07)&~~-1.80(11)&~~-1.86(10)& 42  \\
 14 & 23& V~~I  &&~~-2.20(09)& 11&& -2.10    & 1&&~~-2.27(13)&~~-2.22(11)&~~-1.98(11)& 34  \\
  &   & V~~II   &&          &   &&          &  &&~~-2.22(12)&~~-2.22(17)&~~-2.23(18)&  6  \\
 15 & 24& Cr~I  &&~~-2.26(09)& 10&& -2.20(01)& 2&&~~-2.23(09)&~~-2.32(11)&~~-2.10(16)& 32  \\
  &   & Cr~II   &&~~-2.03(12)&  3&& -2.12(14)& 3&&~~-2.02(08)&~~-2.08(18)&~~-2.04(10)& 10  \\
 16 & 25& Mn~I  &&~~-2.57(08)&  8&& -2.81    & 1&&~~-2.37(13)&~~-2.42(09)&~~-2.25(08)& 12  \\
 17 & 26& Fe~~I &&~~-2.03(11)&187&& -2.04(16)&58&&~~-2.09(07)&~~-2.10(07)&~~-1.92(10)&221  \\
  &   & Fe~~II  &&~~-2.04(11)& 23&& -1.99(07)& 5&&~~-2.09(17)&~~-2.02(17)&~~-2.12(17)& 30  \\
 18 & 27& Co~~I &&~~-1.76(11)&  8&& -1.77(11)& 3&&~~-2.05(12)&~~-2.08(12)&~~-1.88(13)& 22  \\
 19 & 28& Ni~~I &&~~-2.07(08)& 50&& -2.12(11)&15&&~~-2.07(08)&~~-2.13(08)&~~-1.99(07)& 80  \\
 20 & 29& Cu~~I &&~~-2.87(05)&  2&& -2.88(01)& 2&&~~-2.82    &~~-2.82    &~~-2.64    &  1  \\
 21 & 30& Zn~~I &&~~-1.83   &  1&&          &  &&~~-1.82(03)&~~-1.87(04)&~~-1.90(04)&  2  \\
 22 & 38& Sr~~~I&&~~-2.23   &  1&& -2.02    & 1&&~~-2.33    &~~-2.31    &~~-2.12    &  1  \\
  &   & Sr~~~II &&          &   && -2.15    & 1&&~~-2.22    &~~-2.26    &~~-2.36    &  1  \\
 23 & 39& Y~~~II&&~~-2.12(10)& 10&& -2.22(15)& 5&&~~-2.10(04)&~~-2.12(05)&~~-2.15(05)& 13  \\
 24 & 40& Zr~~I &&~~-2.23   &  1&& -2.12    & 1&&~~-2.04(09)&~~-1.94(01)&~~-1.81(10)&  2  \\
  &   & Zr~~II  &&          &   && -2.01    & 1&&~~-2.09(16)&~~-2.13(21)&~~-2.14(18)&  4  \\
 25 & 42& Mo~I    &&          &   && -2.22    & 1&&~~-2.47    &~~-2.47    &~~-2.34    &  1  \\
 26 & 56& Ba~II   &&~~-1.86(04)&  2&& -2.10    & 1&&~~-2.09(12)&~~-2.52(21)&~~-2.52(20)&  3  \\
 27 & 57& La~II   &&~~-1.89(05)&  6&& -1.94(03)& 3&&~~-1.90(06)&~~-1.88(09)&~~-1.90(12)& 12  \\
 28 & 58& Ce~II   &&~~-1.93(12)& 13&& -1.95(15)&14&&~~-1.92(09)&~~-1.90(10)&~~-1.91(09)& 33  \\
 29 & 59& Pr~II   &&~~-1.51(13)&  5&& -1.56(08)& 3&&~~-1.56(14)&~~-1.53(14)&~~-1.55(15)& 11  \\
 30 & 60& Nd~II   &&~~-1.60(08)& 25&& -1.62(11)&12&&~~-1.67(10)&~~-1.64(09)&~~-1.66(09)& 51  \\
 31 & 62& Sm~II   &&~~-1.54(11)&  6&& -1.64(05)& 3&&~~-1.59(08)&~~-1.56(06)&~~-1.58(06)& 12  \\
 32 & 63& Eu~II   &&~~-1.58   &  1&& -1.57(09)& 2&&~~-1.54(13)&~~-1.46(18)&~~-1.47(20)&  6  \\
 33 & 64& Gd~II   &&          &   && -1.55(09)& 3&&~~-1.60(10)&~~-1.57(12)&~~-1.60(12)&  5  \\
 34 & 66& Dy~II   &&~~-1.55   &  1&& -1.25(11)& 6&&~~-1.34(15)&~~-1.34(12)&~~-1.38(10)&  7  \\
 35 & 68& Er~~II  &&~~-1.35   &  1&& -1.38    & 1&&~~-1.41(04)&~~-1.36(06)&~~-1.37(04)&  2  \\
 36 & 69& Tm~II   &&          &   &&          &  &&~~-1.15    &~~-1.23    &~~-1.24    &  1  \\
 37 & 72& Hf~~II  &&          &   &&          &  &&~~-1.23(26)&~~-1.12(15)&~~-1.14(14)&  2  \\
 38 & 74& W~~I    &&          &   &&          &  &&~~-1.69(17)&~~-1.80(19)&~~-1.62(01)&  2  \\
 39 & 76& Os~~I   &&          &   &&          &  &&~~-1.23(13)&~~-1.33(13)&~~-1.14(13)&  4  \\
 40 & 77& Ir~~~I  &&          &   &&          &  &&~~-1.18    &~~-1.12    &~~-1.10    &  1  \\
 41 & 82& Pb~~I   &&          &   &&          &  &&~~-1.85    &~~-1.87    &~~-1.60    &  1  \\
 42 & 90& Th~~II  &&          &   &&          &  &&~~-1.21(12)&~~-1.18(15)&~~-1.19(14)&  7  \\
  &   & Th~II$^2$    &&     &   &&          &  &&~~-1.27(11)&~~-1.23(13)&~~-1.23(14)&  5 \\
 43 & 92& U~~~II  &&          &   &&          &  &&$<$-1.45(30)&$<$-1.45(30)& $<$-1.45(30)&  8  \\
  &   & U~~II$^2$    &&     &   &&          &  &&$<$-1.55(19)&$<$-1.55(19)&$<$-1.55(19)&  5  \\
%   &   &         &&          &   &&          &  &&          &          &          &     \\
 \hline
 \multicolumn{9}{l} {\rm{$^1$ adopted from molecular lines }} \\
 \multicolumn{9}{l} {\rm{$^2$ Only lines with Nilsson et al. (2002a,b) oscillator strengths}} \\
 \end{tabular}
%\normalsize
% }
 \end{table*}

  Table~1  and  2  contain  data on  lines  for  iron and other chemical
  elements.
%  respectively.
%  More  precisely,  ,
  Table~1 provides the  ionization stage, equivalent width,
  oscillator  strength, energy of lower  level and derived iron abundance
  for  each  iron wavelength considered.
  For  the  other elements, Table~2 gives  in  addition to the line data
  (identification,  wavelength,  oscillator strength,  the energy of lower
  level),  the abundance determination  (relative abundance with respect
  to  the  solar  system  value, and  absolute  values  of the abundance
  calculated  from  this  line using the  spectrum  of \hd and the solar
  spectrum), the levels of blending of the line in the synthetic spectra
  of \hd and of the Sun, the depths of the line in the synthetic spectra
  of  \hd  and of the Sun, and  the abundances in the atmosphere of \hd,
  calculated  with two different atmosphere models, namely for a surface
  gravity  increased  by  0.2~dex,  and  for  an  effective  temperature
  increased by 100 \K. % respectively.

 % Figure 4
 \begin{figure}
 \resizebox{\hsize}{!}{\includegraphics{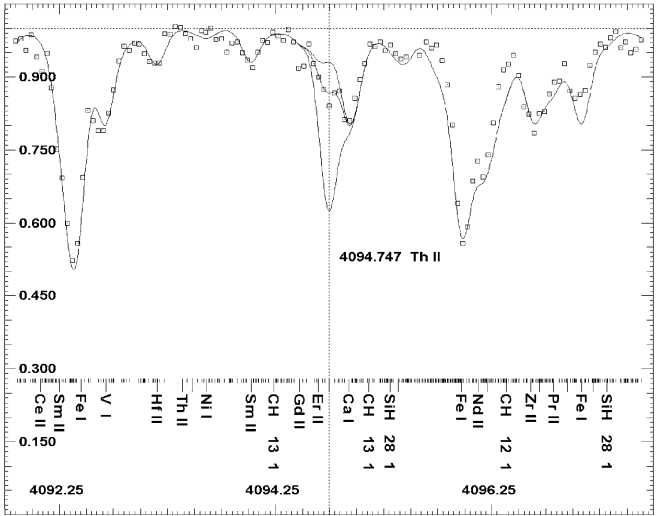}}
 %\vskip 1.5cm
 \caption[]
 {Same as Fig. 3 in the vicinity of the Th II 4094.747 \AA~ line.  }
 \label{fig4}
 \end{figure}

 % Figure 5
 \begin{figure}
 \resizebox{\hsize}{!}{\includegraphics{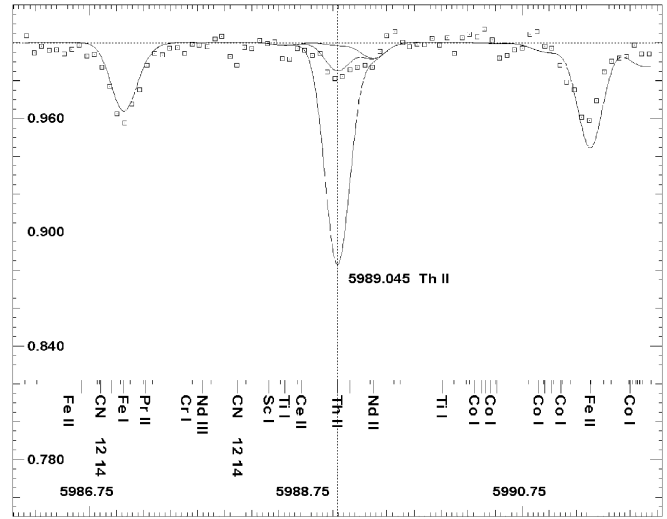}}
\vskip 0.2cm
 \caption[]
 {Same as Fig. 3 in the vicinity of the Th II 5989.045 \AA~ line.  }
 \label{fig5}
 \end{figure}

 % Figure 6
 \begin{figure}
 \resizebox{\hsize}{!}{\includegraphics{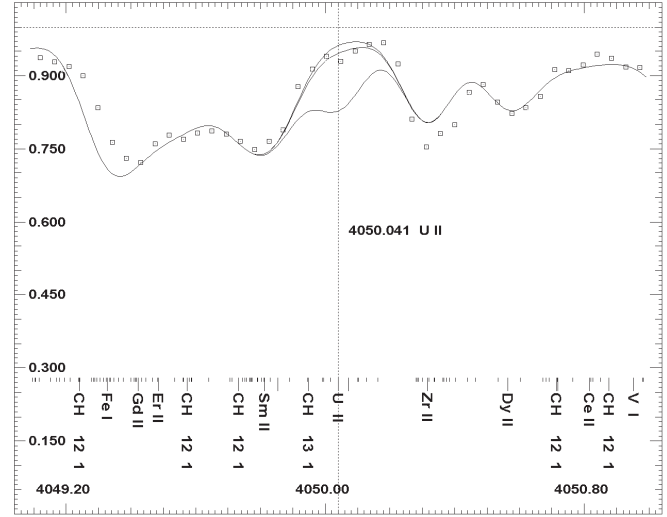}}
 %\vskip 1.5cm
 \caption[]
 {Same as Fig. 1 in
  the  vicinity of the U II  4050.041 \AA~ line. The different synthetic
  spectra  correspond to a uranium abundance  of -2.52, -2.02, and -1.72
  (in the scale log N(H)=12).}
 \label{fig6}
 \end{figure}

  In Table~3, the mean elemental abundances in the atmosphere of \hd are
  given with respect to the solar atmosphere value. For all investigated
  elements,   Table~3   includes  information   on  the  atomic  number,
  designation   of  the  ionization  stage,  as  well  as  the  relative
  abundances  (the  last digits in  brackets correspond to the estimated
  errors)  and  the  number of lines  determined  in  Paper~I and in the
  present  study.  Mean abundances are calculated  with  the best set of
  atmosphere parameters, with a surface gravity increased by 0.2~dex and
  with  an  effective  temperature  increased  by  100~\K.  The absolute
  abundances  are  available in the electronic  version  of the table. A
  comparison  of  our abundances with  previously  published data can be
  found in Paper~I. The abundances of S and Pb can be found here for the
  first  time. The abundances of Mo, Er, Tm,  Hf, W, Os, Ir, Th, and U first
  determined  by  Yushchenko  et al. (2002)  are  updated in the present
  study.

  In  Table~4  we show the differential  and absolute abundances of iron
  and  some  of the key elements  used to trace back the nucleosynthesis
  and   corresponding   cosmochronometry  discussed   below.  The  errors
  correspond  to the standard deviations  of abundances derived from the
  individual  lines of the element. For  uranium and thorium, only lines
  with new oscillator strengths are considered.

 \begin{table}
   \caption{Abundance  of  iron and of  some  elements determined in the
  present  work.  The columns provide,  respectively, the atomic number,
  the  element symbol, the number of  lines analyzed, the mean abundance
  in  the atmosphere of \hd with  respect to solar photosphere (Grevesse
  \& Sauval 1998) and the abundance in the scale logN(H)=12.}
 \begin{tabular}{cl r cr }
 \hline
 Z &  Ident.  &  N   &   $\Delta$log N  &      logN~~~        \\
 \hline
 26  &   Fe I   & 221  &  -2.09+0.07      &     5.41$\pm$0.07  \\
     &   Fe II  & 30   &  -2.09+0.17      &     5.41$\pm$0.17  \\
 63  &   Eu II  & 6    &  -1.54+0.13      &    -1.03$\pm$0.13  \\
 76  &   Os I   & 4    &  -1.23+0.12      &     0.22$\pm$0.12  \\
 77  &   Ir I   & 1    &                  &     0.17$\pm$0.20  \\
 82  &   Pb I   & 1    &  -1.85+0.20      &     0.10$\pm$0.20  \\
 90  &   Th II  & 5    &                  &    -1.18$\pm$0.11  \\
 92  &   U II   & 5    &                  & $<$-2.02$\pm$0.20  \\
 \hline
 \end{tabular}
 \label{Table1}
 \end{table}

 % Figure 7
 \begin{figure}
 \resizebox{\hsize}{!}{\includegraphics{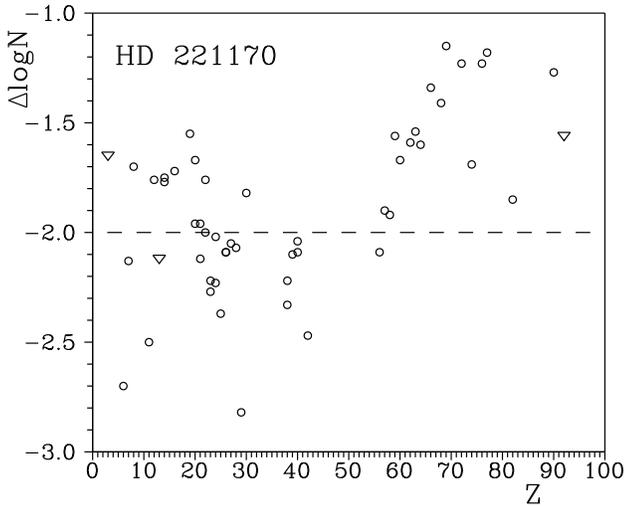}}
 %\vskip 1.5cm
 \caption[]
 {
The abundances of chemical elements and ions in the atmosphere of \hd
with respect to the solar atmosphere value.
Triangles correspond to upper limits for Li, Al, U.
}
 \label{fig7}
 \end{figure}

  Concerning  the  detection of carbon  and nitrogen, their atomic lines
  are  hardly  detectable,  so that  their  abundance  determinations are
  usually  based  on  molecular lines.  We  calculated several synthetic
  spectra  in  the  whole  observed  region  using  different  C  and  O
  abundances  and found abundances close to  -2.7 and -2.13 with respect
  to the Sun. This result is in good agreement with Sneden et al. (1986)
  and  Craft  et  al.  (1992), so that  it  is  adopted in the abundance
  determination  of the other elements.  Special attention was also paid
  to  the determination of the isotopic  carbon ratio which can strongly
  influence  the final results. To estimate the $^{12}$C/$^{13}$C ratio,
  several   synthetic   spectra  were   calculated   and  compared  with
  observation.  We finally obtained the  value of $^{12}$C/$^{13}$C=6 in
  agreement with the determination of Sneden et al. (1986).

  The  wavelengths  of several $^{13}$CH  lines were changed according to
  Johnson  \&  Bolte (2001). The corresponding  list of spectral lines in
  the  vicinity  of the Th~II 4019.128~\AA~  line  was considered in our
  calculations. Note that our result concerning the thorium abundance is
  based  on  5  lines with Nilsson  et  al.  (2002) oscillator strengths.
  Similar  abundances  are obtained with  all  these lines. We therefore
  recommend them for future abundance determinations in other stars.

\section{Discussion}

  About  half of the stable nuclei heavier than iron in the Universe are
  synthesized  by  the rapid neutron-capture  process, also known as the
  r-process.  The  r-process is believed  to  occur predominantly in the
  latest  stages  of evolution of massive  stars  (heavier than about 10
  solar  masses)  during their supernova  explosion. Fig.~8 compares \hd
  abundances  with  the  solar content  in  r-process nuclei. It clearly
  shows  that for elements heavier than Ba (Z=56), the \hd surface abundance
  pattern  is  very similar to the  solar one. Globally, it confirms the
  previous  observation  of ultra-metal-poor stars  (Westin et al. 2000;
  Cayrel et al. 2001; Hill et al. 2002; Cowan et al. 2002; Sneden et al.
  1998, 2003) that already in the early age of the Galaxy, the r-process
  was  operational and quite unique in  its production of, at least, the
  Z=56  to  78 elements all along the  life  of the Galaxy. However, the
  universality  of the r-process for  the production of elements Z$>$56,
  including  the Pb-peak elements and the  actinides, still remains to be
  confirmed.  Such  a  universality has  deep  implications including in
  particular  the invariance of the  relative r-nuclidic abundances with
  respect  to galactic chemical evolution effects and the possibility to
  develop  a  stellar chronometry based on  the  actinide content of the
  metal-poor stars and consequently to estimate a lower limit to the age
  of the Galaxy.

  Although   for   most  of  the   elements  with  Z=56-78  the  present
  observations seem to comfirm this conclusion, the specific cases of Eu
  and  Hf observed in \hd show  the first indication of a non-negligible
  deviation with respect to the solar pattern (Fig. 8). Such a deviation
  remains rather small, but, as shown by Goriely \& Arnould (1997), large
  deviations are not expected in this mass region because of the nuclear
  correlations   inherent  to  the  nuclear  aspects  of  the  r-process
  nucleosynthesis.
% {\bf
  It was shown in the previous section that taking
  into account the hyperfine and isotopic structure of Hf lines could reduce the
  deviation but not to the extent to become compatible with the solar pattern.
  In the case of Eu, the abundance determination includes the splitting effects  in
  both \hd and the Sun.
%  }

  In  this  comparison  with  the  solar  r-abundance
  distribution,  another  interesting  feature  of  \hd  is  its high Pb
  surface abundance. Such an unusually high abundance has been
  already observed
  in  the  previous  ultra-metal-poor  r-process-enriched  stars  except
  CS31082-001.  Lead  can be efficiently  produced by the r-process, but
  also  by  the  slow neutron-capture  process,  known as the s-process,
  during the AGB phase of low-metallicity stars with masses in the range
  0.8-8  solar  masses (Van Eck et al.  2001). Such a Pb s-enrichment is
  however   systematically   accompanied   by   a   simultaneous  carbon
  enrichment.  For this reason, the significant  Pb enrichment of a star
  like  CS22892-052  characterized  by a  large  C/O$>$1  ratio could be
  explained  by    additional  C and  Pb  s-process  enrichment from a
  companion  star.  But, in the specific  case of \hd, its low C/O=0.025
  ratio   characteristic  of  heavy  mass  stars  tend  to
  argue against  this
  possibility.  The  high Pb content of  \hd provides the second strong
  hint  that  the r-process might not  be universal in its production of
  the elements above Ba.

 \begin{table}
 \caption{Comparison of abundance ratios	 (in log scale)
       in Th-rich stars. The second column gives the metallicity
       in terms of [Fe/H].
%   Errors affecting the Th/Pb ratio is of the order
%   of 0.3 dex and for the Th/Eu ratio 0.1 dex.
   }
\begin{tabular}{ l c c r l }
 \hline
   Star            & [Fe/H] & Th/Eu &  Th/Pb       &  Th/U          \\
 \hline
 HD221170          & -2.1   & -0.15 &    -1.28     & $>$0.84        \\
%
%CS31082-001       & -2.9   & -0.19 & $>$-0.69     & ~~0.81$\pm$0.1 \\  % old value
 CS31082-001$^a$   & -2.9   & -0.22 & $>$-0.78     & ~~0.94         \\  % Hill et al. 2002
 CS22892-052$^b$   & -3.1   & -0.62 &    -1.62     & $>$0.73        \\
 HD115444$^c$      & -3.0   & -0.60 &    -1.8$^d$  & $>$0.40        \\
 BD+17$^o$3248$^e$ & -2.1   & -0.51 & $>$-1.48     & $>$0.82        \\
 \hline
 \end{tabular}
\\
~   $^a$Hill   et al. (2002). \\
~   $^b$Sneden et al. (2003). \\
~   $^c$Westin et al. (2000) for all ratios except Th/Pb. \\
~   $^d$Th: Westin et al. (2000), Pb: Sneden et al. (1998). \\
~   $^e$Cowan et al.  (2002). \\
 \label{Table2}
 \end{table}

  Finally, the present observation provides an accurate determination of
  the surface Th as well as an upper limit of U abundances in \hd. These
  actinides have been used extensively in recent years to estimate the
  age  of the oldest metal-poor stars of the Galaxy and in so doing, a lower
  limit  to  the age of our Galaxy.  The Th/Eu, Th/Pb and Th/U abundance
  ratios  in  \hd are compared in  Table 5 to the corresponding ratios
  determined  in the other four  r-process-rich metal-poor stars.
%(Burris  et  al.  2000;  Westin et al. 2000;  Cayrel  et al. 2001; Hill et al. 
%  2002; Cowan et al. 2002;  Sneden  et al. 2003). 
  These  ratios  are of particular interest
  since they compare the Th abundance to light species (represented here
  by Eu), Pb (believed to be produced in the same ``environment'' as Th,
  see e.g Goriely \& Clerbaux, 1999) and U, a neighboring element. Based
  on  the  universality  of the  r-process,  a  relatively reliable Th/U
  chronometry  can  be built. An absolute  determination of the Th and U
  abundances  has been achieved for  CS31082-001 only, leading to
  an  age  estimate  of  13$\pm$4 Gyr  (Hill et al. 2002; Goriely \&
  Arnould  2001).  From Table 5, it can  be  deduced from the Th/U ratio
  that,  within  the  universality assumption, \hd  is  older or at most
  2~Gyr younger than CS31082-001.

  The  \hd  Th/Eu  ratio is seen to  be  the highest among the different
  stars,   which  is   compatible  with  its  relatively  higher
  metallicity   [Fe/H].  However,  assuming   the  universality  of  the
  r-process,  i.e  the  same original  actinide  production in all these
  stars  (relative  to the $Z=56-78$  r-elements), this would imply that
  \hd  is  younger  than  CS22892-052 by  about  22$\pm$4  Gyr, but also
  younger  by  17$\pm$4  Gyr  than  BD+17$^o$3248,  a  star  of  similar
  metallicity.  This would imply that BD+17$^o$3248 would be at least 28
  Gyr  old,  an  observation  that is  difficult  to  reconcile with the
  traditional  Big-Bang  cosmology
%  {\bf
  or the WMAP observation
%  }.
 It is also  seen  that
as far as the
  Th/Eu  ratio is concerned, \hd  is compatible with CS31082-001, clearly
  confirming
%{\bf
% for the first time
% }
  that the Th-rich star CS31082-001  is not
  a  rare  exceptional  case.
%{\bf
  This conclusion was also drawn by Honda et al. (2004) who determined at the
surface of CS30306-132 a ratio of [Th/Eu]=-0.10 even higher than for CS31082-001 or
     HD221170.
%  }
  The Th/Pb  ratios  shown  in Table 5 again
  confirms  the previous conclusion that there is a large scatter in the
  r-abundances for the heaviest elements. The Th/Pb chronometry assuming
  the  universality  of the r-process would  lead to the conclusion that
  HD115444  is older than CS31082-001  (of similar metallicity) by about
  48$\pm$14 Gyr.

\section{Conclusion}

  We determined the abundances of 43 chemical elements (including Th and
  U)  in  the atmosphere of the  bright  halo star HD221170. The present
  spectroscopic   observation  indicates  that   the  r-process  is  not
  universal  or  more  exactly  the  astrophysical  site(s)  hosting the
  r-process  do(es)  not  always  lead  to  a  unique relative abundance
  distribution  for  the  bulk Ba to Hg  elements,
% on  a first side,
  the Pb-peak  elements,
% on a second side,
  and the actinides.
% on a third side.
  Although  the  previous  observations  of  CS22892-052,  HD115444  and
  BD+17$^o$3248  suggested  the universal feature  of the r-process, the
  present  \hd  analysis  confirms  the  non-universal  trend originally
  suggested by the CS31082-001 observation. This new finding rules out
  the  use of  thorium chronometry to  estimate the age of the oldest
  stars in the Galaxy.

 % Figure 8
 \begin{figure}
 \resizebox{\hsize}{!}{\includegraphics{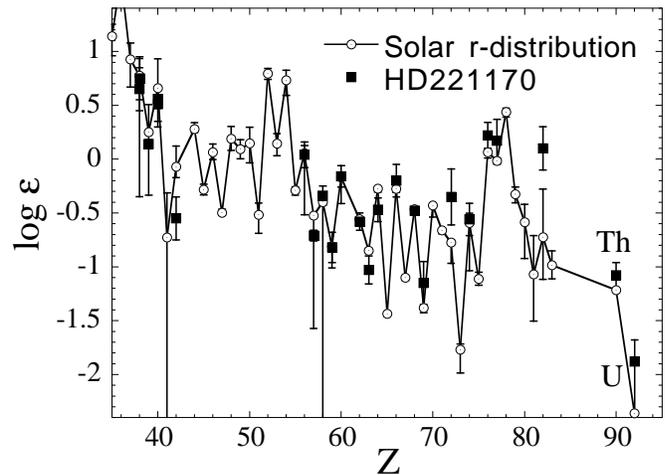}}
% \vskip 0.5cm
 \caption[]
 {
    Comparison  of  the  surface abundances  in  HD221170 with the solar
  system  r-abundance distribution scaled to the observed Er abundance.
  The   solar   system  abundances   corresponds   to  the  photospheric
  determination  of  Grevesse \& Sauval  (1998) with the solar r-process
  distribution estimated by Goriely (1999).
 }
 \label{fig8}
 \end{figure}

\begin{acknowledgements}
  We  would like to thank to L.  Delbouille and G. Roland for sending us
  the  Liege Solar Atlas. We use data from NASA ADS, SIMBAD, CADC, VALD,
  NIST,  and DREAM databases and we  thank the teams and administrations
  of   these  projects.  Work  by  AY   and  YK  was  supported  by  the
  Astrophysical  Research Center for the  Structure and Evolution of the
  Cosmos  (ARCSEC)  of Korea Science  and Engineering Foundation (KOSEF)
  through the Science Research Center (SRC) program.
  Work by VG and CK was supported
  by research funds of  Chonbuk National University, Korea.
  S.G. is FNRS Research Associate.
\end{acknowledgements}

%\clearpage


\begin{thebibliography}{}
%
\bibitem [2002]{22}  Biemont, J., Palmeri, P., \&  Quinet, P.  2002,
	      Database of rare earths at Mons University //
	      http://www.umh.ac.be/\~ astro/dream.html
%
%
%\bibitem [1957]{23} Burbidge, E.M., Burbidge, G.R.,  Fawler, W.A., \& Hoyle, F.
%	      1957, Reviews Modern Physics, 29,  547
%
\bibitem [2000]{01} Burris, D., Pilachowski, C.A., Armandroff, T.E.,
		    Sneden, C., Cowan, J.J., \& Roe, H.
		     2000,  ApJ, 544, 302
%
\bibitem [2001]{02} Cayrel, R.,  Hill, V., Beers, T. C., Barbuy, B.,
		    Spite, M., Spite, F., Plez, B., Andersen, J.,
		    Bonifacio, P., Francois, P., Molaro, P., Nordstrom, B., \&
		    Primas, F.
		     2001, Nature, 409, 691.
%
\bibitem [2002]{03} Cowan, J.J., Sneden, C., Burles, S., Ivans, I.I.,
		    Beers, T.C., Truran, J.W., Lawler, J.E., Primas, F.,
		    Fuller, G.M., Pfeiffer, B., \& Kratz, K.-L.
		      2002, ApJ, 572, 861
%
\bibitem [1973]{25}  Delbouille, L., Roland, G., \&  Neven, L.   1973:
              Photometric  Atlas  of  the  Solar  Spectrum  from
              $\lambda$3000  to $\lambda$10000,
              Li\'{e}ge:  Institut d'Astrophysique de l'Universite' de Liege.
%
\bibitem [1993]{03a} Francois, P., Spite, M., \& Spite, F.,
                   1993, A\&A, 274, 821
%
\bibitem [1992]{26}  Galazutdinov, G.A. 1992, SAO RAS Preprint No.92
%
%\bibitem [2003]{05}  Galazutdinov, G., Musaev, F., Bondar, A., \& Krelovsky, J.
%                     2003, MNRAS, 345, 365
%
\bibitem [1988]{05ab} Gilroy, K.K., Sneden, C., Pilachowski, C., Cowan, J.,
                    1988,  ApJ, 327, 298
%
\bibitem [1999]{05a}  Gopka, V.F., Yushchenko, A.V., Shavrina, A.V., \&
                    Perekhod, A.V., 1999,
                    Kinematika Fiz. Nebesn. Tel., 15, 447
%
\bibitem [2001]{06}  Gopka, V., Yushchenko, A., Mishenina, T., \& Kovtyukh,
		     2001, Odessa Astron. Publ, 14, 237
%
\bibitem [2004]{G04} Gopka, V.F., Yushchenko, A.V., Mishenina, T.V.,
		     Kim, C., Musaev, F.A., \& Bondar, A.V. 2004,
                     Astronomy Reports, 48, 577
                     (Paper I)
%
\bibitem [1997]{07}  Goriely,  S., \& Arnould, M. 1997, A\&A, 322, L29
%
\bibitem [1999]{08b}  Goriely,	S. 1999, A\&A, 342, 881

\bibitem [1999]{08}  Goriely,  S. \& Clerbaux B. 1999, A\&A, 346, 798

\bibitem [2001]{10}  Goriely,  S., \&  Arnould, M. 2001, A\&A, 379, 1113


\bibitem [1998]{15z} Grevesse, N., \& Sauval, A.J.,
		     1998, Space Science Reviews 85, 161
%
\bibitem [1998]{09}  Grevesse, N., \& Sauval, A.J.
		     1999, A\&A, 347, 348
%
\bibitem [2002]{0hc}  Hill, V., Plez, B., Cayrel, R., Beers, T.C.,
                    Nordstrom, B., Andersen, J., Spite, M., Spite, F.,
                    Barbuy, B., Bonifacio, P., Depagne, E.,
                    Francois, P., Primas, F.
                    2002, A\&A, 387, 560
%
\bibitem [2003]{04ac} Honda, S., Aoki, W., Kajino, T., Ando, H.,
                    Beers, T.C., Izumiura, H., Sadakane, K.,
                    Takada-Hidai, M.,
                    2004, ApJ, 607, 474
%
\bibitem [2003]{04cc} Ivarsson, S., Andersen, J., Nordstrom, B., Dai, X.,
                      Johansson, S., Lundberg, H., Nilsson, H., Hill, V.,
                      Lundqvist, M., \& Wyart, J. F.
                      2003, A\&A, 409, 1141
%
\bibitem [1995]{11aa} Johnson, J., \& Bolte, M., 2001, ApJ, 554, 888
%
\bibitem [1995]{11}  Kurucz, R.L. 1995, ASP Conf. Ser., 81, 583
%
%\bibitem [2004]{12} Kurucz, R.L. 2004,   http://www.kurucz.harvard.edu
%
\bibitem [1992]{25a} Kraft, R.P., Sneden, C., Langer, G.E, \& Prosser, C.F.
                     1992, AJ, 104, 645
%
%\bibitem [1992]{12a} Morrel, Kallander, \& Bucher, 1992,
                     A\&A, 259, 543
%
%\bibitem [2000]{28} Morton, D.C. 2000, ApJS, 130, 403
%
\bibitem [1999]{13}  Musaev, F., Galazutdinov, G.,  Sergeev, A.,
		     Karpov, N., \& Pod'yachev Y.
		     1999, Kinematics Phys. Select. Bodies, 15, 216
%
\bibitem [2002]{14}  Nilsson,	H.,  Ivarsson,	S.,   Johansson,   S.,	\&
		     Lundberg,	H.
		     2002a, A\&A, 381, 1090
%
\bibitem [2002]{15}  Nilsson, H., Zhang, Z., Lundberg, H., Johansson, S., \&
		     Nordstrom, B.
		     2002b, A\&A, 382, 368
%
\bibitem [1986]{16a} Shavrina, A.V., Polosukhina, N.S.,
       Pavlenko, Ya.V., Yushchenko, A.V., Quinet, P., Hack,
       M., North, P., Gopka, V.F., \'Zverko, J., \v{Z}i\v{z}\v{n}ovsk\'{y}, J.,
       \&  Veles, A.
       2003, A\&A,  409, 707
%
\bibitem [2003]{16b} Sneden,  C.,  Cowan, J.J., Burris, D.L., Truran, J.W.
                     1998, ApJ, 496, 235
%
\bibitem [2003]{16}  Sneden,  C.,  Cowan, J.J., Lawler, J.E.,
		     Ivans, I.I., Burles, S., Beers, T.C., Primas, F.,
		     Hill, V., Truran, J.W., Fuller, G.M.,
		     Pfeiffer, B., \& Kratz, K.-L.
		     2003, ApJ, 591, 936
%
\bibitem [2001]{17}  Van  Eck,	S., Goriely, S.,  Jorissen,  A., \& Plez, B.
		     2001, Nature, 412, 793
%
\bibitem [1963]{18a}  Wallerstein,  G., Greenstein, J.L., \& Parker, R.,
                      1963, ApJ, 137, 280
%
\bibitem [2000]{18}  Westin,  J.,  Sneden,  C.,  Gustafsson,   B.,  \&
		     Cowan,  J.
		     2000, ApJ, 530, 783
%
\bibitem [1998]{19} Yushchenko, A.V. 1998,
		     Proc. of  the  29th conf. of variable star  research,
		     Brno, Czech Republic, November 5-9, 201
%
\bibitem [1998]{19a} Yushchenko, A.V., Gopka, V.F., 1994,
                      Astronomy  Letters, 20, 453
%
\bibitem [2002]{JKAS} Yushchenko, A., Gopka, V., Kim, C., Khokhlova, V.,
	Shavrina, A., Musaev, F., Galazutdinov, G., Pavlenko, Y.,
	Mishenina, T., Polosukhina, N., \&   North, P. 2002, Journal Korean
	Astron. Soc., 35, 209
%
\bibitem [2004]{20}  Yushchenko, A.V., Gopka, V. F., Kim, C., Liang, Y. C.,
                     Musaev, F. A., \& Galazutdinov, G. A.
		     2004, A\&A, 413, 1105
%
\end{thebibliography}
\end{document}